\documentclass[acmsmall]{acmart}
\AtBeginDocument{%
  }

\setcopyright{cc}
\setcctype{by}
\acmDOI{10.1145/3832192}
\acmYear{2026}
\acmJournal{PACMSE}
\acmVolume{3}
\acmNumber{ISSTA}
\acmArticle{ISSTA101}
\acmMonth{10}
\acmSubmissionID{issta26main-p903-p}
\received{2026-01-30}
\received[accepted]{2026-04-16}

\usepackage{amsmath,amsfonts}
\usepackage{algorithmic}
\usepackage{graphicx}
\usepackage{textcomp}
\usepackage[ruled,vlined,linesnumbered]{algorithm2e}
\usepackage{multirow}
\usepackage{adjustbox}
\usepackage{pifont}
\usepackage{threeparttable}
\usepackage{booktabs}
\usepackage{bbding}
\usepackage{tablefootnote}
\usepackage{hyperref}
\usepackage{tcolorbox}
\usepackage{enumitem}
\usepackage{enumerate}
\usepackage{subfig}
\usepackage{url}
\usepackage{soul}
\usepackage{balance}
\usepackage{makecell}
\usepackage{color}
\usepackage{listings}
\usepackage{float}
\usepackage{ulem}
\hypersetup{
colorlinks=true,
urlcolor=black,
linkcolor=black,
citecolor=black
}
\usepackage{wrapfig}

\newcommand\yc[1]{\textcolor{black}{#1}}

\newcommand{\tech}{SeGa}
\newcommand{\tool}{SeGa}

\newcommand{\deploynum}{6}
\newcommand{\fixed}{16}

\newcommand{\bugtype}{business logic bug}

\newcommand{\nobase}{\tool$_{w/o\text{ }base}$}
\newcommand{\nofunc}{\tool$_{w/o\text{ }func}$}
\newcommand{\noscenario}{\tool$_{w/o\text{ }scenario}$}
\newcommand{\norepair}{\tool$_{w/o\text{ }repair}$}

\newcommand{\tester}{CHATTESTER}
\newcommand{\sym}{SymPrompt}
\newcommand{\hits}{HITS}
\newcommand{\rat}{RATester}

\begin{document}

\title{Uncovering Business Logic Bugs via Semantics-Driven Unit Test Generation (Experience Paper)}

\author{Chen Yang}
\orcid{0000-0003-0759-940X}
\affiliation{%
  \institution{Tianjin University}
  \city{Tianjin}
  \country{China}}
\email{yangchenyc@tju.edu.cn}

\author{Junjie Chen}
\orcid{0000-0003-3056-9962}
\authornote{Corresponding author}
\affiliation{%
  \institution{Tianjin University}
  \city{Tianjin}
  \country{China}}
\email{junjiechen@tju.edu.cn}

\begin{abstract}
Business logic bugs violate intended business semantics and are particularly prevalent in enterprise software. 
Yet most existing unit test generation techniques are code-centric, making such bugs difficult to expose. 
We present \tech{}, a semantics-driven unit test generation technique for uncovering business logic bugs. 
\tech{} constructs a semantic knowledge base from product requirement documents, represented as a set of functionality entries that group related requirements under a common business intent. Given a focal method, \tech{} retrieves the relevant functionality entries and derives fine-grained business scenarios with explicit preconditions, triggering actions, expected outcomes, and semantic constraints to guide LLM-based test generation. We evaluate \tech{} on four industrial Go projects containing 60 real-world business logic bugs. \tech{} detects 22$\sim$25 more bugs than four state-of-the-art LLM-based techniques and improves precision by 26.9\%$\sim$34.6\%. Deployment across \deploynum{} production repositories further uncovers \fixed{} previously unknown business logic bugs that were confirmed and fixed by developers, demonstrating \tool{}’s practical value. From our industrial study, we summarize a series of lessons and suggestions for practical use and future research.

\end{abstract}



\maketitle

\lstdefinestyle{codestyle}{
    numbers=left,                   %
    numberstyle=\tiny\color{gray},   %
    basicstyle=\ttfamily\scriptsize,
    lineskip=0.8pt,                   %
    keywordstyle=\color{magenta},    %
    commentstyle=\color{violet!60!gray}, %
    stringstyle=\color{blue},         %
    backgroundcolor=\color{white}, %
    showstringspaces=false,          %
    xleftmargin=1.5em,               %
    xrightmargin=1.5em,              %
    frame=single,                    %
    breaklines=true,                 %
    moredelim=**[is][\color{red!50}]{`}{`}, %
    escapeinside=``,                 %
    columns==fullflexible,
    float=htbp,
}

\lstdefinestyle{textstyle}{
    numbers=none,                   %
    numberstyle=\tiny\color{gray},   %
    basicstyle=\ttfamily\scriptsize,
    lineskip=0.8pt,                   %
    keywordstyle=\color{magenta},    %
    commentstyle=\color{violet!60!gray}, %
    stringstyle=\color{blue},         %
    backgroundcolor=\color{white}, %
    showstringspaces=false,          %
    xleftmargin=1.5em,               %
    xrightmargin=1.5em,              %
    frame=single,                    %
    breaklines=true,                 %
    moredelim=**[is][\color{red!50}]{`}{`}, %
    escapeinside=``,                 %
    columns==fullflexible,
}

\section{Introduction}

Automatic unit test generation plays a crucial role in modern software testing by enabling systematic and scalable validation of program behavior. 
A large body of prior work has studied Search-Based Software Testing (SBST), in which tests are generated or evolved to maximize structural coverage criteria such as statement, branch, or path coverage~\cite{evosuite,randoop,pynguin}. 
More recently, advances in Large Language Models (LLMs) have enabled a new class of test generation techniques that produce unit tests directly from source code and its surrounding context, often achieving coverage comparable to or even exceeding that of traditional SBST approaches~\cite{chattester,chatunitest,symprompt,wang2024hits,telpa,rat}.

Despite their differing technical foundations, both lines of work are fundamentally code-centric: they rely primarily on information extracted from the code itself to guide test generation.
As a result, they are effective at uncovering defects that manifest in code structure or execution behavior, such as null-pointer dereferences~\cite{lin2021graph}, invalid operations~\cite{unitcon}, and type errors~\cite{rted}. 
However, by emphasizing how the code executes rather than what the software is intended to do, these techniques largely overlook intended business semantics defined by domain requirements.
This limitation becomes particularly problematic in enterprise software, where correctness is governed by complex business policies, multi-step workflows, and legacy rules that are rarely made explicit in source code. 
Consequently, bugs arising from mismatches between implemented code and intended business semantics (referred to in this work as \textbf{business logic bugs}) remain difficult to detect using existing code-centric unit test generation techniques.

Incorporating business semantics beyond source code into unit test generation is therefore a promising direction for addressing such bugs.
While both search-based and learning-based techniques could, in principle, benefit from richer semantic guidance, LLMs provide a particularly natural interface for consuming and reasoning over heterogeneous semantic information, which is often expressed in textual artifacts such as requirement documents.
Accordingly, this work focuses on LLM-based unit test generation. However, enriching LLMs with external business semantics also introduces several challenges:

\textbf{Challenge 1: Business semantics extraction from informal specifications.}
The first challenge lies in reliably extracting high-quality business semantics from informal requirement artifacts.
In practice, business semantics are primarily documented in product requirement documents (PRDs), which contain requirements about expected behavior, valid operational workflows, domain constraints, and implicit assumptions.
However, PRDs are designed for human communication rather than automated analysis.
They are often noisy and weakly structured, with core business semantics interleaved with background information, implementation notes, historical context, and change logs.
Moreover, relevant semantic information is frequently scattered across multiple sections, expressed at different levels of abstraction, or embedded in non-textual formats such as tables or figures.
These characteristics make it difficult to distill precise and actionable semantic representations that can serve as reliable guidance for unit test generation.

\textbf{Challenge 2: Effective utilization of extracted semantics for unit test generation.}
The second challenge concerns how to effectively utilize the extracted business semantics to guide the testing of a specific focal method.
A single PRD typically covers a wide range of functionalities, only a small subset of which are relevant to the method under test.
At the same time, the focal method often contains substantial implementation-specific logic (such as framework interactions, auxiliary computations, and defensive checks) that is orthogonal to the underlying business intent.
This gap between document-level semantics and method-level code behaviors complicates the alignment of relevant semantics.
Without effective utilization mechanisms, extracted semantics may be either too coarse-grained or misaligned with the focal method, limiting their effectiveness in generating realistic, targeted, and semantically meaningful unit tests.

To address the two challenges, we propose \tech{} (\textbf{Se}mantics-driven unit test \textbf{G}ener\textbf{a}tion), a semantics-driven unit test generation approach that enhances LLM-based techniques by systematically integrating code context with business semantics extracted from requirement documents.
\textbf{First}, \textit{\tech{} constructs a semantic knowledge base from raw requirement documents.}
It leverages a multimodal LLM to convert tables and figures into textual form, filters out content unrelated to business semantics, and extracts functionalities that specify what the software is intended to do.
Each extracted functionality is encoded as a structured domain-specific language (DSL) entry, which reduces ambiguity and supports efficient retrieval and downstream semantic reasoning during test generation.
\textbf{Second}, \textit{given a focal method, \tech{} retrieves method-relevant functionality entries from the semantic knowledge base to guide unit test generation.}
Specifically, a semantic reasoning agent analyzes the focal method together with its necessary surrounding code context and summarizes the method’s business intent as a concise natural-language description.
This summary is then used as a query to retrieve relevant functionality entries.
Because a retrieved functionality entry may aggregate multiple requirements, only some of which apply to the focal method, the semantic reasoning agent derives fine-grained business scenarios that isolate the requirement(s) relevant to the focal method and encode them in a scenario DSL.
Each scenario is represented using explicit \textbf{preconditions}, \textbf{triggering actions}, \textbf{expected outcomes}, and \textbf{constraints}, thereby providing precise and focused semantic context for subsequent unit test generation.
\textbf{Finally}, \textit{a test generation agent jointly reasons over the semantic context and the code context to generate unit tests}. 
To ensure practical applicability in real-world projects with strict build and dependency constraints, \tech{} incorporates a standalone compilation-repair component that automatically fixes compilation issues in generated tests, thereby improving their executability.

We evaluated \tech{} on four large-scale, production Go projects developed and maintained by a leading international IT company (i.e., ByteDance). These projects are actively used in industrial settings and collectively involve 60 real-world \bugtype{}s that were discovered and fixed during normal development and maintenance.
The selected projects span multiple business lines and system types, including inter-service communication components (e.g., RPC-based microservices), external-facing interfaces (e.g., backend API implementations), and core production services that support critical business workflows. As a result, they exhibit diverse implementation styles, dependency structures, and operational constraints typical of industrial software systems.
The evaluated \bugtype{}s reflect a broad range of business logic bugs encountered in practice. These include security-related issues (e.g., inconsistencies in user authentication logic and improper handling of sensitive credentials), concurrency errors (e.g., deadlocks arising from multi-service interactions), data-flow and integration bugs (e.g., incorrect propagation or transformation of inputs across upstream and downstream services), and functional correctness bugs (e.g., partially or incorrectly implemented business functionalities). Together, these bugs provide a realistic and representative basis for assessing \tech{} in industrial environments.
We compared \tech{} against several state-of-the-art LLM-based unit test generation techniques, including \tester{}~\cite{chattester}, \sym{}~\cite{symprompt}, \hits{}~\cite{wang2024hits}, and \rat{}~\cite{rat}.
Specifically, \tech{} detects 29 \bugtype{}s, whereas \tester{}, \sym{}, \hits{}, and \rat{} detect only 7, 7, 6, and 4 bugs, respectively. 
Overall, \tech{} uncovered 22$\sim$25 more \bugtype{}s than existing techniques.
Moreover, \tech{} attained a precision of 0.73, compared to 0.54, 0.54, 0.55, and 0.57 for \tester{}, \sym{}, \hits{}, and \rat{}, respectively. 
These results indicate that \tech{} not only detects substantially more business logic bugs, but also produces fewer false positives in industrial settings.

Furthermore, we deployed \tech{} on \deploynum{} production repositories spanning multiple business lines within the company, including core product logic, backend services, and infrastructure components. 
These repositories are developed and maintained by independent engineering teams with diverse design conventions and development practices, reflecting realistic industrial heterogeneity.
\textbf{Using \tech{}, we uncovered \fixed{} previously unknown \bugtype{}s, all of which were confirmed and subsequently fixed by the developers. }
Feedback from the participating developer teams highlighted that the generated tests were not only accurate in exposing faulty behavior, but also helped surface subtle requirements that were previously implicit.
Together, these results demonstrate the practical effectiveness of \tech{} in uncovering real-world \bugtype{}s and its value as a testing aid in industrial development environments.

We further analyzed the false positives produced by \tech{}. 
Some false positives arose when the agent hypothesized implausible failure scenarios (e.g., incorrect behavior of downstream services or databases) that are unlikely to occur in deployed production software. 
Other false positives were caused by ambiguities, inconsistencies, or missing details in the requirement documents, which led the model to infer behaviors not intended by the developers. 
In addition, in certain cases, \tech{} flagged behaviors that, while potentially risky under general best-practice guidelines, were intentionally implemented according to team-specific engineering conventions, where developers rely on compensating mechanisms or contextual practices that are not visible to the model.
These observations highlight that high-quality requirement documentation remains critical: even sophisticated LLMs cannot reliably compensate for unclear or incomplete specifications. 
Moreover, accurate detection of \bugtype{}s requires not only knowledge of explicit requirements, but also an understanding of team-specific practices. We further discuss these in Section~\ref{sec:rq1}.

In summary, our contributions are as follows:
\begin{itemize}
    \item We design \tech{}, a semantics-driven approach that systematically integrates business semantics extracted from requirement documents into LLM-based unit test generation, enabling the detection of \bugtype{}s beyond the reach of code-centric approaches.

    \item We conduct a large-scale industrial evaluation at a global IT company, demonstrating that \tech{} detects substantially more real-world \bugtype{}s than state-of-the-art LLM-based techniques and has been successfully deployed across \deploynum{} production repositories.

    \item We distill practical lessons learned from deploying semantics-guided unit test generation in industrial settings, providing actionable insights for both researchers and practitioners.
    
\end{itemize}

\section{Background and Motivation}

\subsection{Terminology}
\label{sec:background}
In this paper, business semantics are sourced from requirement documents (e.g., PRDs).
These documents contain (1) requirements that specify expected software behavior and (2) irrelevant content that does not describe expected behavior and distracts semantic analysis.
Specifically, we introduce four terms used in this paper to concretize business semantics.
\begin{itemize}
    \item \textbf{Requirement}: 
    a requirement is the atomic unit of business semantics extracted from a PRD, specifying an expected system behavior at a high level (distinguished from irrelevant content such as backgrounds in PRDs).

    \item \textbf{Functionality}: 
    a  high-level specification that groups together related requirements under a common business intent.
    For example, the PRD excerpt in Figure~\ref{fig:motivation} contains a functionality named \textit{Item Operation Management}, which aggregates the requirements of managing user operations on items (e.g., edit entry behavior, UI/navigation conditions, permission checks, and user-facing feedback).
    A functionality therefore captures what the system is expected to enforce in general, while leaving case-specific details (e.g., concrete item states, triggering actions, and expected outputs) unspecified.

     \item \textbf{Business intent}: 
     a business intent is the common purpose that unifies the requirements within a functionality, describing what the functionality is intended to implement (e.g., storage item management), and serving as a key anchor for retrieval that enables \tech{} to identify relevant functionalities for a focal method.
    
    \item \textbf{Business scenario}:
    a concrete scenario derived from a functionality.
    Each scenario isolates one requirement within the functionality and makes it explicit by stating preconditions, a triggering action, and expected outcomes, together with the semantic constraints that must hold for that scenario.
    For instance, from \textit{Item Operation Management}, the scenario \textit{Editing forbidden under legacy mode} concretizes requirement~(2) by specifying \textbf{preconditions} (the item is in legacy mode), a \textbf{triggering action} (invoke edit from the main view), and \textbf{expected outcomes} (editing is disabled and a reason is returned).
    Each scenario also carries the \textbf{semantic constraints} (e.g., the prohibition under legacy mode and the corresponding input/state constraints), which can guide unit test construction.
\end{itemize}
Conceptually, a functionality groups multiple requirements under a shared business intent, and each requirement can be instantiated into a concrete business scenario. 

\subsection{Motivating Example}

\begin{figure*}[t]
  \centering
  \includegraphics[width=0.85\linewidth]{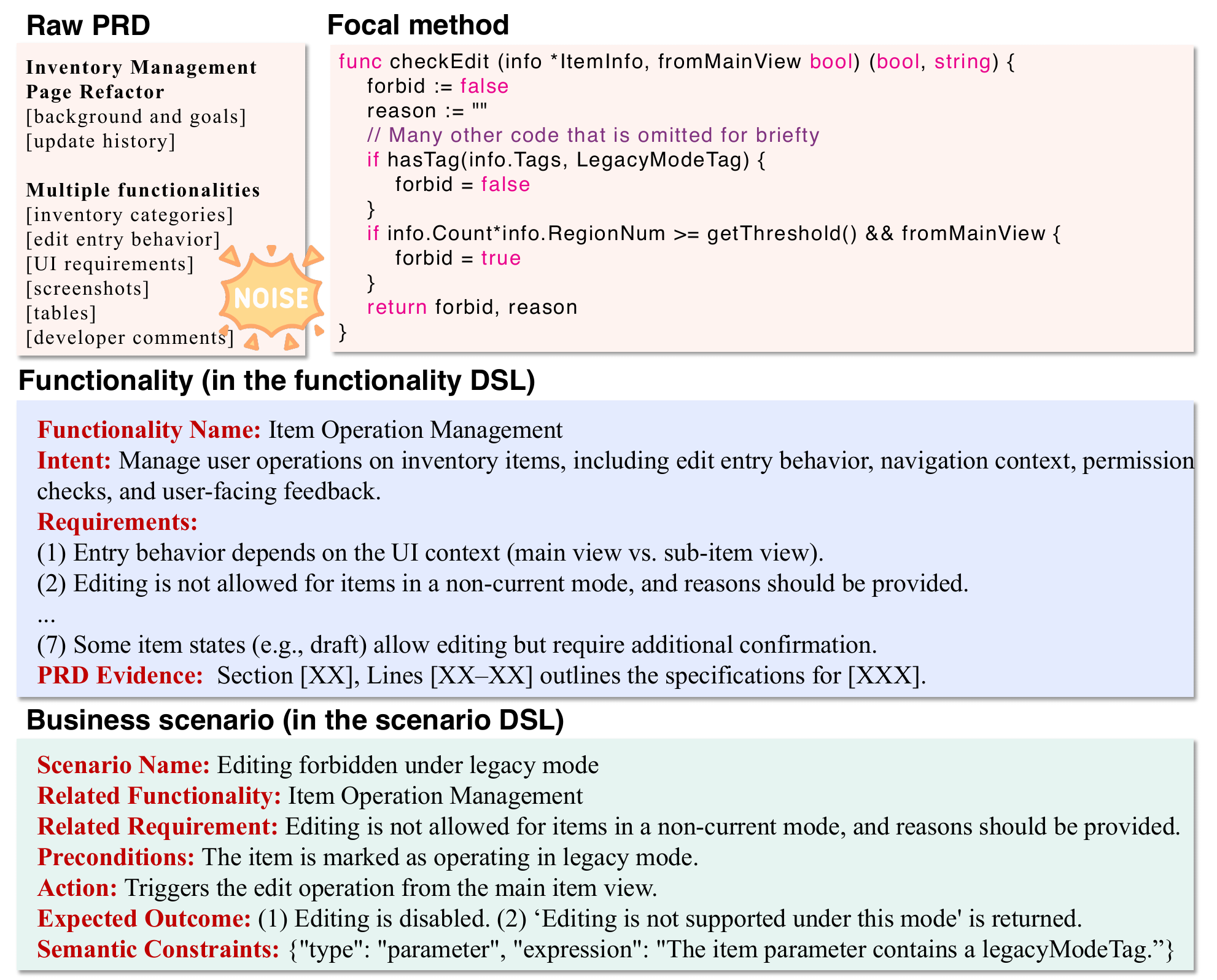}
  \caption{Motivating Example}
  \label{fig:motivation}
\vspace{-1em}
\end{figure*}

Figure~\ref{fig:motivation} illustrates a motivating example adapted from an enterprise software.\footnote{The code and requirements shown are desensitized and simplified versions of internal software used in production.}
The focal method determines whether an editing operation on an item should be forbidden.
According to the PRD, items operating in \texttt{legacy-mode} should not allow editing.
However, in the illustrated code, when an item is marked as \texttt{legacy-mode}, the method explicitly allows the edit operation by setting \texttt{forbid} to \texttt{false}.
As a result, the method is syntactically valid, yet violates the requirement.
Since this violation is neither encoded in the control flow nor reflected in the method signature, it cannot be inferred from source code alone.
Consequently, existing code-centric unit test generation techniques, including \tester{}, \sym{}, \hits{}, and \rat{}, primarily follow the implemented logic and fail to expose this \bugtype{}.

A natural idea is to provide the PRD as additional context for test generation.
In practice, this is ineffective.
In the software underlying Figure~\ref{fig:motivation}, the `legacy-mode-edit-forbidden' requirement is embedded within several pages of PRD that describe many unrelated contents.
When provided with the full document, LLM-based generators either overlook the specific requirement that editing must be forbidden under \texttt{legacy-mode}, or are distracted by the noise.
As a result, even with PRDs as context, state-of-the-art techniques including \tester{}, \sym{}, \hits{}, and \rat{} still fail to expose the bug.
\textit{
This motivates abstracting raw PRDs into a structured representation based on functionalities, which reduces noise by organizing dispersed requirements into coherent units that can be retrieved and used for reasoning about expected behavior.
}

However, providing all functionalities extracted from a PRD as context remains ineffective. 
A typical PRD contains many functionalities, only a small subset of which are relevant to a given focal method.
In Figure~\ref{fig:motivation}, the PRD contains over 100 functionalities, whereas the focal method is concerned only with the \textit{Item Operation Management} functionality.
When all functionalities are provided as context, LLM-based generators (e.g., \tester{}, \sym{}, \hits{}, and \rat{}) are easily distracted by irrelevant information and generate tests that either miss the bug or trigger spurious failures.
\textit{
This motivates a semantic retrieval mechanism that identifies the specific functionality (or small set of functionalities) relevant to the focal method.
}

Even after retrieving the relevant functionality/functionalities, the retrieved information is often still too coarse for unit test generation. 
This is because a functionality aggregates multiple requirements, and each requirement is typically stated at a high level, while a focal method usually realizes only a subset of the requirements within that functionality (e.g., requirement~(2) in Figure~\ref{fig:motivation}). 
Providing the entire functionality as guidance can therefore blur which requirement(s) the method is intended to realize and leave details underspecified, misleading test generation. 
For example, when guided by the retrieved \textit{Item Operation Management} functionality, \rat{} fails to construct valid inputs and an executable invocation path, producing a non-compiling test that still misses the bug.
\textit{This motivates deriving specific business scenario(s) from the retrieved functionality. 
Each business scenario isolates one requirement that is relevant to the focal method and makes it actionable for unit test construction by specifying concrete preconditions, a triggering action, and expected outcomes, together with the semantic constraints that must hold for that case.}

Guided by the scenario(s), \tech{} can generate semantically meaningful unit tests.
Figure~\ref{fig:motivation} shows the retrieved functionality \textit{Item Operation Management} and the derived scenario \textit{Editing forbidden under legacy mode}.
For clarity, the figure shows only the functionality relevant to the focal method and the scenario that triggers the bug.
Using this semantic guidance, \tech{} successfully generates a unit test that exposes the bug missed by existing code-centric techniques.

\section{Approach}
\label{sec:approach}

\begin{figure*}[t]
  \centering
  \includegraphics[width=\linewidth]{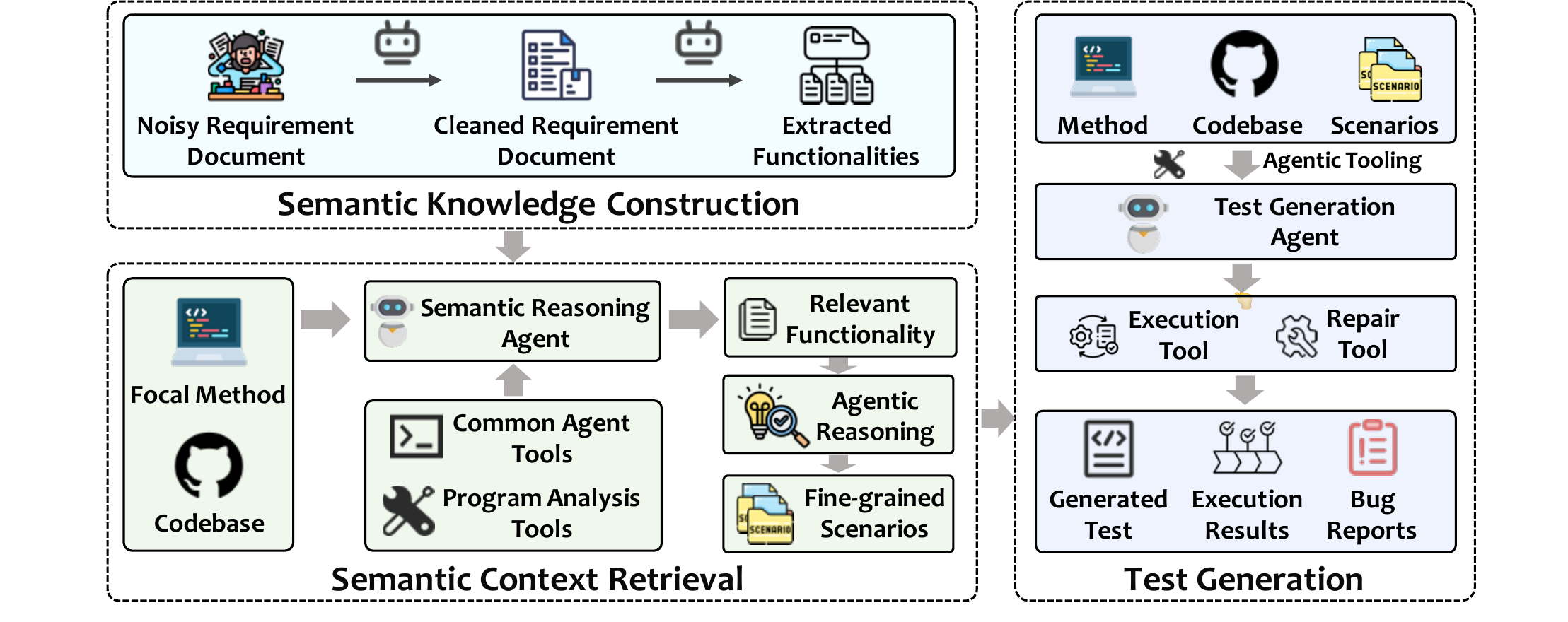}
  \caption{Overview of \tech{}}
  \label{fig:overview}
\vspace{-1em}
\end{figure*}

Figure~\ref{fig:overview} shows an overview of \tech{}. 
To address the limitations of code-centric unit test generation in detecting \bugtype{}s, \tech{} treats product requirement documents (PRDs) as an external source of business semantics. 
However, since PRDs are noisy, weakly structured, and often expressed at a granularity that does not align with method-level code behavior, making it difficult for LLMs to distill precise and actionable test guidance, \tech{} processes them through a three-stage workflow.
First, \textbf{semantic knowledge construction} cleans and normalizes PRDs, extracting functionality entries stored in a structured functionality DSL. 
Second, \textbf{semantic context retrieval} leverages a semantic reasoning agent to summarize the focal method’s intended behavior and retrieve relevant functionality entries. 
Since a functionality entry may include multiple requirements, the agent derives business scenarios in a scenario DSL, each of which concretizes one requirement relevant to the focal method.
Finally, \textbf{semantics-guided test generation} uses a test generation agent to reason over the derived business scenarios and the code context, and generates unit tests.

\subsection{Semantic Knowledge Construction}
\label{sec:semantic_knowledge_construction}
Raw requirement documents are often noisy and weakly structured, and thus directly using them could be ineffective.
Therefore, \tech{} constructs a semantic knowledge base from raw requirement documents.
Specifically, \tech{} (1) normalizes and cleans documents to remove irrelevant content and unify formats, and (2) decomposes the text into structured functionalities that capture coherent units of intended requirements.
These functionality entries form the knowledge base used for subsequent semantic context retrieval and semantics-guided test generation.

\subsubsection{Requirement Document Normalization}

In practice, requirement documents contain a large amount of content that does not contribute to business semantics, such as background descriptions and historical context.
Such content obscures the requirements that specify expected software behavior and interferes with subsequent semantic analysis.
In addition, requirements are not always expressed as plain text.
Some requirements are specified in non-textual forms such as tables or figures.
If left unprocessed, these representations prevent complete recovery of business semantics from the document.
To address these challenges, \tech{} performs requirement document normalization to produce a text-only representation that preserves business semantics while removing irrelevant content.
First, non-textual elements are interpreted and converted into natural-language descriptions using a multimodal LLM (i.e., SeedCode), and reinserted at their original positions to preserve contextual continuity.
Next, \tech{} uses the LLM to identify and remove document fragments that do not describe expected software behavior, such as high-level background explanations.
The resulting normalized document preserves requirements that specify expected software behavior and are needed for accurate functionality construction.

\subsubsection{Functionality Extraction}
Even after normalization, a requirement document typically contains many requirements describing software behavior at different levels of abstraction, often interleaved across sections.
Using the document directly as guidance is therefore ineffective, because requirements of varying scope and granularity make it difficult for subsequent analysis.
To address this, \tech{} decomposes each normalized document into a set of functionalities, where each functionality groups together related requirements under a common business intent, as defined in Section~\ref{sec:background}.
This decomposition step produces explicit functionality entries that serve as the basic semantic units for further processing.
Specifically, \tech{} represents each functionality entry using a domain-specific language (DSL). Compared to free-form text, the DSL provides a standardized schema that reduces ambiguity and enables consistent downstream processing. 
Building on prior work in software requirements engineering~\cite{dsl1,dsl2}, we design a tailored functionality DSL for our task. Each functionality entry includes the following fields:
\begin{figure}[H]
    \vspace{-4mm}
    \centering
    \includegraphics[width=1.0\linewidth]{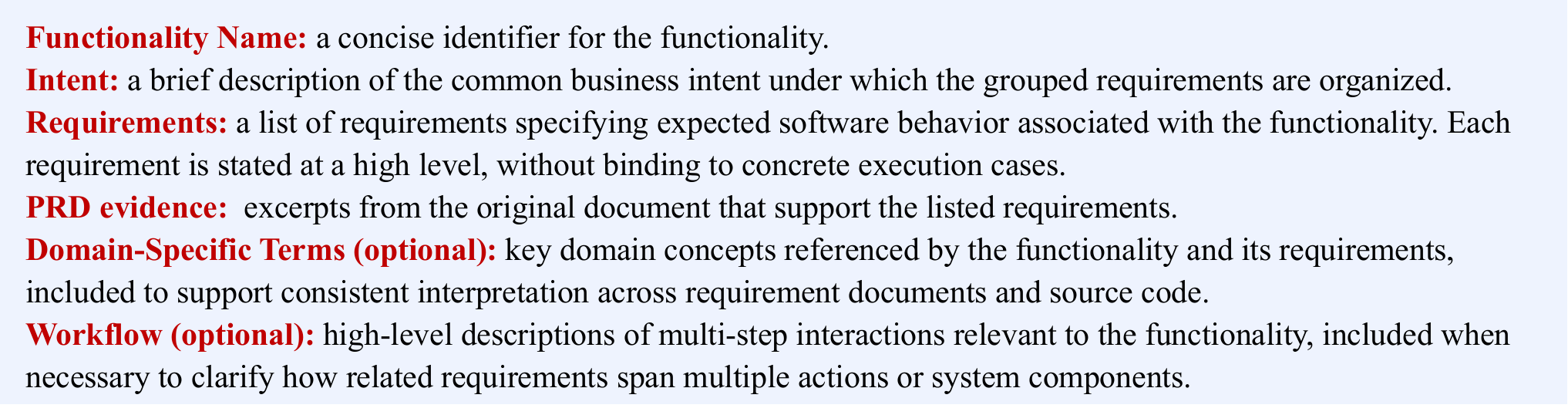}
    \label{fig:functionality_dsl}
    \vspace{-10mm}
\end{figure}
\noindent

To construct functionality entries in the functionality DSL, \tech{} employs an LLM-based extraction and structuring procedure.
The LLM is provided with (1) the normalized requirement document, (2) the formal definition of the functionality DSL, and (3) task instructions that guide the model to identify functionalities that capture common business intents and the corresponding requirements originally described in the document.
These requirements are then summarized and organized under each functionality.
Through this process, \tech{} produces a standardized DSL representation for each functionality, which serves as the foundation for subsequent semantic context retrieval, and ultimately supports semantics-guided unit test generation.
Figure~\ref{fig:motivation} shows an example functionality entry represented using this DSL.

\subsection{Semantic Context Retrieval}
\label{sec:semantic_context}
To translate the constructed semantic knowledge into actionable guidance for a focal method, \tech{} first retrieves the functionality (or a small set of closely related functionalities) relevant to the method from the knowledge base.
This is challenging because functionalities are expressed as high-level groupings of requirements in natural language, whereas a method’s intent is encoded in source code and often interleaved with auxiliary or framework-related logic, making direct matching unreliable.
To address this gap, \tech{} employs a semantic reasoning agent that performs hierarchical reasoning over code and natural language.
Starting from the focal method, the agent inspects the method and its necessary surrounding code to summarize the method’s intent in natural language, and uses this summary to retrieve the relevant functionality entry/entries from the knowledge base.
However, retrieved functionalities are still too coarse for unit test construction. By definition, a functionality aggregates multiple requirements under a common business intent, while a focal method typically realizes only a subset of them. \tech{} therefore derives business scenarios from the retrieved functionality entry/entries. Each business scenario isolates one requirement applicable to the focal method and makes it explicit by specifying concrete preconditions, a triggering action, and expected outcomes, together with the semantic constraints. These business scenarios serve as the semantic context for the subsequent test generation stage.

\subsubsection{Relevant Functionality Retrieval}
Given a focal method, the semantic reasoning agent analyzes the method together with the necessary surrounding code to infer what the method is intended to do.
To support this process, the agent uses two sets of tools.
\textbf{Common agent tools} provide basic interaction with the repository, including \texttt{View} (list directories or view file contents) and \texttt{Bash} for executing shell commands such as \texttt{find} and \texttt{grep}.
These tools allow the agent to locate relevant code.
\textbf{Program-analysis tools} complement them by extracting program information that is hard to obtain reliably from text search alone, such as resolved call relations and type/constant bindings:
\begin{itemize}
    \item \textbf{const\_finder}: locates constant definitions referenced by a method.
    \item \textbf{var\_type\_finder}: resolves variable definitions and their associated types.
    \item \textbf{callchain\_finder}: identifies the callee functions invoked by a method.
    \item \textbf{func\_info\_finder}: extracts a method’s signature, parameters, and return types.
    \item \textbf{struct\_finder}: retrieves the definitions of referenced struct types and their members.
\end{itemize}
The program-analysis tools are built on AST analysis and exposed to the agent as executable commands.
They return targeted information (e.g., call relations and resolved types), enabling the agent to infer method intent without scanning large amounts of code.

Based on the collected information, the agent summarizes the method’s intended behavior as a short natural-language description.
The summary focuses on what the method is supposed to achieve, the key domain concepts it involves, and the observable outcomes, while ignoring auxiliary logic such as logging, error wrapping, and framework-related checks.
The agent then uses this summary as a query to retrieve relevant functionality entries from the semantic knowledge base.
Since entries are stored in a structured DSL, the agent matches the summary against fields such as the intent, requirements, and domain terms.
This retrieval step narrows the search space to a small set of relevant functionalities, which are then refined into business scenarios for test generation.

\subsubsection{Scenario Derivation}
While retrieved functionality entry/entries group requirements under common business intents, deriving guidance for testing a specific focal method requires identifying which requirement applies to the method and how it can be exercised in code.
This cannot be determined during the knowledge base construction, which relies only on requirement documents, but instead requires joint reasoning over the retrieved functionality and the focal method’s implementation.
Therefore, at this stage, the semantic reasoning agent proceeds to derive business scenarios as the semantic context for the subsequent test generation stage.
Concretely, the agent jointly analyzes the focal method’s code context together with the retrieved functionality to select the requirement(s) applicable to the method and refine them into concrete scenarios.
Each business scenario corresponds to one requirement from the retrieved functionality entry/entries that applies to the focal method, and makes that requirement explicit by specifying preconditions, a triggering action, and expected outcomes, together with the semantic constraints that must hold for the case. Each business scenario is represented using a scenario DSL.
\begin{figure}[H]
    \vspace{-3mm}
    \centering
    \includegraphics[width=1.0\linewidth]{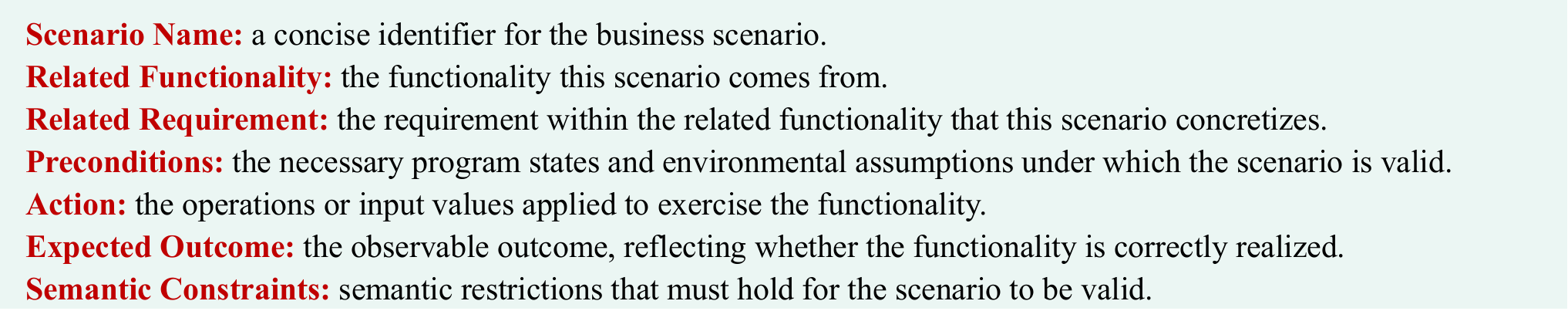}
    \label{fig:scenario_dsl}
    \vspace{-10mm}
\end{figure}
\noindent
Within a scenario entry, semantic constraints are an integral component that restrict how the scenario can be validly exercised.
Constraints may take different categories (i.e., restrictions on input parameters, required program states, and domain-specific invariants).
To support downstream reasoning and test generation, \tech{} records each constraint in a structured JSON format with two fields: \texttt{type} (the constraint category) and \texttt{description} (a natural-language statement of the restriction).
Figure~\ref{fig:motivation} illustrates an example of a business scenario represented using this DSL.

\subsection{Semantics-Guided Unit Test Generation}
\label{sec:test_generation}
Although business scenarios provide semantic context through their preconditions, action, expected outcomes, and semantic constraints, generating an executable test still requires code-specific details. 
In particular, constructing valid inputs, setting up dependencies (e.g., mocks), and invoking correct APIs depend on the focal method’s implementation and its surrounding code context. 
Therefore, \tech{} generates tests by jointly leveraging the semantic context and the code context, so that the resulting tests are guided by semantics and executable in the target project.

\subsubsection{Test Generation}
The process starts by collecting bootstrap code context needed to generate an executable unit test for the focal method. 
Specifically, \tech{} extracts (1) the method signature (receiver, parameters, and return values), (2) parameter/return types (including referenced structs and interfaces), (3) referenced constants/variables with resolved types, and (4) direct callees and their signatures, which often reveal required dependencies and initialization. 
Beyond this initial context, the generation agent acquires additional repository information on demand via standard tools (e.g., \texttt{View}, \texttt{Create}, \texttt{Insert}, \texttt{StrReplace}, \texttt{Bash}) and program-analysis tools.
Guided by the collected code context and the semantic context, the agent generates \yc{one test for each scenario} by mapping preconditions to setup code, the action to a concrete invocation, and expected outcomes to assertions. 
Tests are written to files and executed in the same environment. To make execution reliable and efficient, \tech{} uses a dedicated execution tool that compiles and runs tests in a controlled Docker environment and returns structured feedback (e.g., compile errors, failing tests, stack traces). 
To bound cost, \tech{} limits the agent to at most 50 interactions.

\subsubsection{Compilation Repair}
Prior studies have shown that LLMs are prone to hallucinations, which can lead to invalid unit tests~\cite{rted,ase_empirical}. As a result, even with semantics-guided generation, LLM-produced tests may fail to compile due to issues such as missing imports or incorrect API usage, particularly in languages with strict compilation rules (e.g., Go). 
Repairing these errors through additional test generation agent iterations is costly and may distract the agent from following the business scenarios. 
Therefore, \tech{} performs compilation repair in a separate tool that runs independently of the generation agent.
When compilation fails, \tech{} collects compiler diagnostics along with the repair context: (1) the generated test code and file path; (2) the focal method and its enclosing package; (3) build and test configuration (e.g., testing framework, dependency declarations, language version); and (4) the symbols referenced in the diagnostics (e.g., unresolved types or functions). 
It then prompts a standalone repair LLM to fix only the reported compilation errors (e.g., adding imports, resolving type mismatches, or adjusting setup code to match framework conventions), while preserving the test behavior implied by the business scenario. 
The loop stops when the test compiles or after three repair attempts.
By separating compilation repair from test generation, \tech{} improves compilation success while keeping the generation process efficient and focused on business-relevant behaviors.

\section{Evaluation Design}
To evaluate \tool{}, we formulate the following research questions (RQs):
\begin{itemize}
    \item \textbf{RQ1: To what extent can \tool{} detect \bugtype{}s compared to state-of-the-art techniques?}
    The primary goal of \tool{} is to generate unit tests that can effectively expose \bugtype{}s. This RQ investigates its ability in detecting such bugs.
    
    \item \textbf{RQ2: How does each main component of \tool{} contribute to its overall effectiveness?}
    This RQ motivates an ablation study to evaluate the individual impact of \tool{}'s core components on its effectiveness.
    
    \item \textbf{RQ3: Can \tool{} uncover previously unknown bugs in enterprise production repositories?}
    This RQ examines the practical applicability of \tool{} by evaluating its ability to discover previously unknown \bugtype{}s in real-world industrial codebases.
\end{itemize}

\subsection{Subjects}
\label{sec:subject}
To answer RQ1 and RQ2, we used four large-scale production Go repositories from a leading international IT company (i.e., ByteDance).\footnote{To preserve confidentiality, we anonymize project identifiers and omit internal product names.}
\yc{
These projects are actively used in industrial settings and collectively involve 60 high-severity real-world \bugtype{}s (rated P0–P1) that were discovered and fixed during normal development and maintenance.
We refer to them as \textit{Subject-1} (\textit{12} bugs), \textit{Subject-2} (\textit{9} bugs), \textit{Subject-3} (\textit{37} bugs), and \textit{Subject-4} (\textit{2} bugs).
}
The repositories cover diverse business lines and system types, including inter-service communication components (e.g., RPC-based microservices), external-facing interfaces (e.g., backend API implementations), and core production services supporting critical business workflows. 
\yc{
Furthermore, they range from \textit{177.2}$\sim$\textit{6,153.2} KLoC and contain \textit{106}$\sim$\textit{1,361} source files across \textit{63}$\sim$\textit{327} packages/modules, with \textit{267}$\sim$\textit{639} third-party dependencies declared in their build configurations. 
}
They therefore reflect typical industrial diversity and complexity in implementation style, dependencies, and operational constraints. 
Each repository is paired with a corresponding PRD.
\yc{
The associated PRDs are also substantial: on average, each PRD contains \textit{72,101} tokens, \textit{64} images, and \textit{14} tables, highlighting the heterogeneity of real-world enterprise requirements.
}
\begin{table}[t]
\renewcommand{\arraystretch}{1}
\caption{Statistics of Subjects}
\centering
\label{tab:subjects}
\resizebox{0.99\linewidth}{!}
{
\begin{threeparttable}
\begin{tabular}{l|ccccccc}
\toprule
\textbf{Subject} & \textbf{Files} & \textbf{KLoC} & \textbf{Packages} & \textbf{Dependencies} & \textbf{Bugs} & \textbf{PRD Tokens} & \textbf{Requirements} \\
\midrule
subject-1 & 313  & 377  & 106 & 562 & 12 & 45{,}975  & 34  \\
subject-2 & 253  & 177  & 84  & 638 & 9  & 114{,}098 & 82  \\
subject-3 & 1361 & 6153 & 327 & 639 & 37 & 115{,}048 & 106 \\
subject-4 & 106  & 505  & 63  & 267 & 2  & 13{,}283  & 8   \\
\bottomrule
\end{tabular}
\end{threeparttable}
}
\vspace{-10pt}
\end{table}

The details of the subjects are shown in Table~\ref{tab:subjects}.
The bugs also cover a broad range of types, including security-related issues (e.g., inconsistencies in authentication logic and improper handling of sensitive credentials), concurrency errors (e.g., deadlocks from multi-service interactions), data-flow and integration bugs (e.g., incorrect propagation or transformation of inputs across services), and functional correctness bugs (e.g., partially or incorrectly implemented functionalities).
\yc{
The buggy focal methods are non-trivial with \textit{4}$\sim$\textit{134} lines of code. On average, each method has \textit{6} branches, invokes \textit{8} direct callees, and references symbols defined in \textit{8} external files (e.g., structs/interfaces, constants, helper functions), indicating substantial cross-file dependencies typical of production services.
}

Each bug is precisely annotated with its location (line ranges) and detailed bug descriptions. 
The corresponding buggy focal method is also reviewed by the developers to ensure that no additional undiscovered bugs are present. These annotations provide the ground truth for determining whether a generated test exposes the intended bug.
We focus on industrial projects because open-source projects rarely provide PRDs of comparable details and complex enterprise business logic needed to evaluate \bugtype{} detection.

To answer RQ3, we applied \tool{} to an additional set of \deploynum{} production repositories within the company. 
These repositories span diverse business lines, including core product logic, backend service, and infrastructure components, and were developed by independent teams following different design conventions and engineering practices. 
By sourcing repositories across multiple projects and teams, we avoid biases introduced by recurring design patterns or shared bug characteristics, which ensures a more representative and challenging evaluation of \tool{} in real-world scenarios.

\subsection{Compared Techniques}
\tool{} targets \bugtype{} detection through unit test generation. Since static tools are generally ineffective for such bugs and no mature SBST tools for Go exist, we compare \tool{} against state-of-the-art LLM-based unit test generation techniques:
\begin{itemize}
\item \textbf{\tester{}}~\cite{chattester}: A typical technique to leverage LLMs for unit test generation by prompting the model with the focal method and its surrounding context.

\item \textbf{\sym{}}~\cite{symprompt}: A path-oriented prompting strategy that asks the LLM to generate tests across diverse execution paths for more comprehensive program exploration.

\item \textbf{\hits{}}~\cite{wang2024hits}: A slice-based approach that decomposes complex methods into smaller code fragments and generates tests for each fragment, which can help reach hard-to-trigger behaviors by simplifying the generation target.

\item \textbf{\rat{}}~\cite{rat}: A repository-aware technique for Go that enhances test generation by injecting global contextual information via the Go language server (\texttt{gopls}).
\end{itemize}

\subsection{Measurements}
\label{sec:metric}
We consider a \bugtype{} to be detected if a generated unit test triggers either an assertion failure or a runtime panic when executed.

\subsubsection{Outcomes on Buggy Methods}
To evaluate bug detection effectiveness, we run each technique on every buggy focal method and execute the generated tests. Following prior work~\cite{toga, rted}, we categorize outcomes of each technique as follows:
\begin{itemize}
\item True Positive for bug detection (TP$_{\textit{bug}}$): 
The technique reports a bug that matches one of the ground-truth bugs in the focal method. A match is determined through manual inspection, by comparing the triggered failure or panic with the annotated bug range and its description.

\item False Positive for bug detection (FP$_{\textit{bug}}$): 
The technique reports a bug that does not match to any ground-truth bug in the focal method.

\item False Negative for bug detection (FN$_{\textit{bug}}$): a bug is not detected by the technique.

\end{itemize}
Since all focal methods in this setting are known to be buggy, there are no true negatives.

\subsubsection{Metric Calculation}
Based on these outcomes, we measured the effectiveness of each technique using the following metrics:

\textbf{Precision} measures the proportion of true bugs among all samples identified as bugs:
$\frac{ \text{TP}_\textit{bug}}{ \text{TP}_\textit{bug} +  \text{FP}_\textit{bug}}$.

\textbf{Recall} measures the proportion of true bugs correctly identified out of all true bugs: 
$\frac{ \text{TP}_\textit{bug}}{ \text{TP}_\textit{bug} +  \text{FN}_\textit{bug}}$.

\textbf{F1-score} is the harmonic mean of Precision and Recall, providing a balanced measure that accounts for both false positives and false negatives:
$\frac{2 \times \text{Precision} \times \text{Recall}}{\text{Precision} + \text{Recall}}$.

\subsection{Implementation and Environment}
We implemented \tool{} in Python and Go. The program-analysis tools are written in Go on top of the built-in AST and packaged as executable binaries for agent invocation. The agents are implemented in Python using LangChain (v1.1.1)~\cite{langchain} and are powered by the SeedCode LLM via API~\cite{seedcode}.
For the baseline techniques \tester{}, \sym{}, and \hits{}, which were originally proposed for Java or Python, we adapted their publicly available implementations to Go and validated their behavior using representative examples, strictly following the approach described in their papers and artifacts.
\rat{} is natively designed for Go. Therefore, we reused its released artifacts and replaced its local LLM calls with API-based calls for consistency.
For fairness, all methods use the same SeedCode model.
For all focal methods in the benchmarks, we configure each technique to generate one test file per focal method, which ensures that the resulting test suites are of comparable scale and enables a fair comparison. 
All generated tests are executed using the projects’ original testing framework, primarily \texttt{Go testing}.
All experiments were conducted on a workstation running Ubuntu 20.04, equipped with a 128-core CPU and 504 GB of RAM.

\section{Results and Analysis}
\label{sec:results}

\subsection{RQ1: Effectiveness on Business Logic Bug Detection}
\label{sec:rq1}
\subsubsection{Process}
We generated and executed tests to observe whether any \bugtype{}s were triggered for each focal method. 
Each technique's output was then mapped to one of several possible outcomes introduced in Section~\ref{sec:metric} based on whether it reported a true \bugtype{}.

\subsubsection{Results}

\begin{table}[t]
\renewcommand{\arraystretch}{1}
\caption{Comparison among \tool{}, \tester{}, \sym{}, \hits{}, and \rat{}}
\centering
\label{tab:rq1}
\resizebox{0.99\linewidth}{!}
{
\begin{threeparttable}
\begin{tabular}{l|ccc|ccc|ccc|ccc}
\toprule
\multicolumn{1}{c|}{\multirow{2}{*}{\textbf{method}}} & \multicolumn{3}{c|}{\textbf{Subject-1}} & \multicolumn{3}{c|}{\textbf{Subject-2}} & \multicolumn{3}{c|}{\textbf{Subject-3}} & \multicolumn{3}{c}{\textbf{Subject-4}} \\
\multicolumn{1}{c|}{} & \multicolumn{1}{c}{\textbf{Precision}} & \multicolumn{1}{c}{\textbf{Recall}} & \multicolumn{1}{c|}{\textbf{F1}} & \multicolumn{1}{c}{\textbf{Precision}} & \multicolumn{1}{c}{\textbf{Recall}} & \multicolumn{1}{c|}{\textbf{F1}} & \multicolumn{1}{c}{\textbf{Precision}} & \multicolumn{1}{c}{\textbf{Recall}} & \multicolumn{1}{c|}{\textbf{F1}} & \multicolumn{1}{c}{\textbf{Precision}} & \multicolumn{1}{c}{\textbf{Recall}} & \multicolumn{1}{c}{\textbf{F1}} \\
\midrule
\tester{} & 0.50 & 0.08 & 0.14 & 0.50 & 0.22 & 0.31 & 0.67 & 0.11 & 0.19 & 0.00 & 0.00 & 0.00 \\
\sym{} & 0.33 & 0.08 & 0.13 & 0.67 & 0.22 & 0.33 & 0.60 & 0.08 & 0.14 & 0.50 & 0.50 & 0.50 \\
\hits{} & 0.33 & 0.08 & 0.13 & 0.50 & 0.11 & 0.18 & 0.60 & 0.08 & 0.14 & 1.00 & 0.50 & 0.67 \\
\rat{} & 0.50 & 0.08 & 0.14 & 0.00 & 0.00 & 0.00 & 0.50 & 0.05 & 0.10 & 1.00 & 0.50 & 0.67 \\
\midrule
\tech{} & \textbf{0.71} & \textbf{0.42} & \textbf{0.53} & \textbf{0.86} & \textbf{0.67} & \textbf{0.75} & \textbf{0.67} & \textbf{0.43} & \textbf{0.52} & \textbf{1.00} & \textbf{1.00} & \textbf{1.00} \\
\bottomrule
\end{tabular}
\end{threeparttable}
}
\end{table}

Table~\ref{tab:rq1} presents the overall results in terms of precision, recall, and F1-score.
From this table, \tech{} consistently outperforms all compared techniques across all projects, achieving the highest F1-score in every case and substantially improving recall on \bugtype{}s while maintaining higher precision.
Specifically, 
aggregated over all projects, \tech{} attains a precision, recall, and F1 of 0.73, 0.48, and 0.58, compared with 0.54, 0.12, and 0.19 for \tester{}, 0.54, 0.12, and 0.19 for \sym{}, 0.55, 0.10, and 0.17 for \hits{}, and 0.57, 0.07, and 0.12 for \rat{}. 
These results show that \tech{} identifies substantially more true \bugtype{}s than existing methods and also achieves stronger precision in most cases, which together lead to consistently higher F1 across all subjects.
We further observe performance differences across projects. All techniques perform worse on Subject-1 and Subject-3 than on the other two projects. 
This is because these two projects involve more complex cross-module business logic and richer state interactions, making it harder to construct bug-triggering inputs. Nevertheless, even under the challenging conditions, \tech{} detects 5 and 16 bugs on Subject-1 and Subject-3, respectively, whereas the strongest baseline, \tester{}, detects only 1 and 4. 
The result demonstrates that \tech{} remains effective even when business semantics are deeply intertwined with implementation details.

\begin{wrapfigure}{r}{6.5cm}
  \vspace{-1em}
  \centering
  \includegraphics[width=0.45\textwidth]{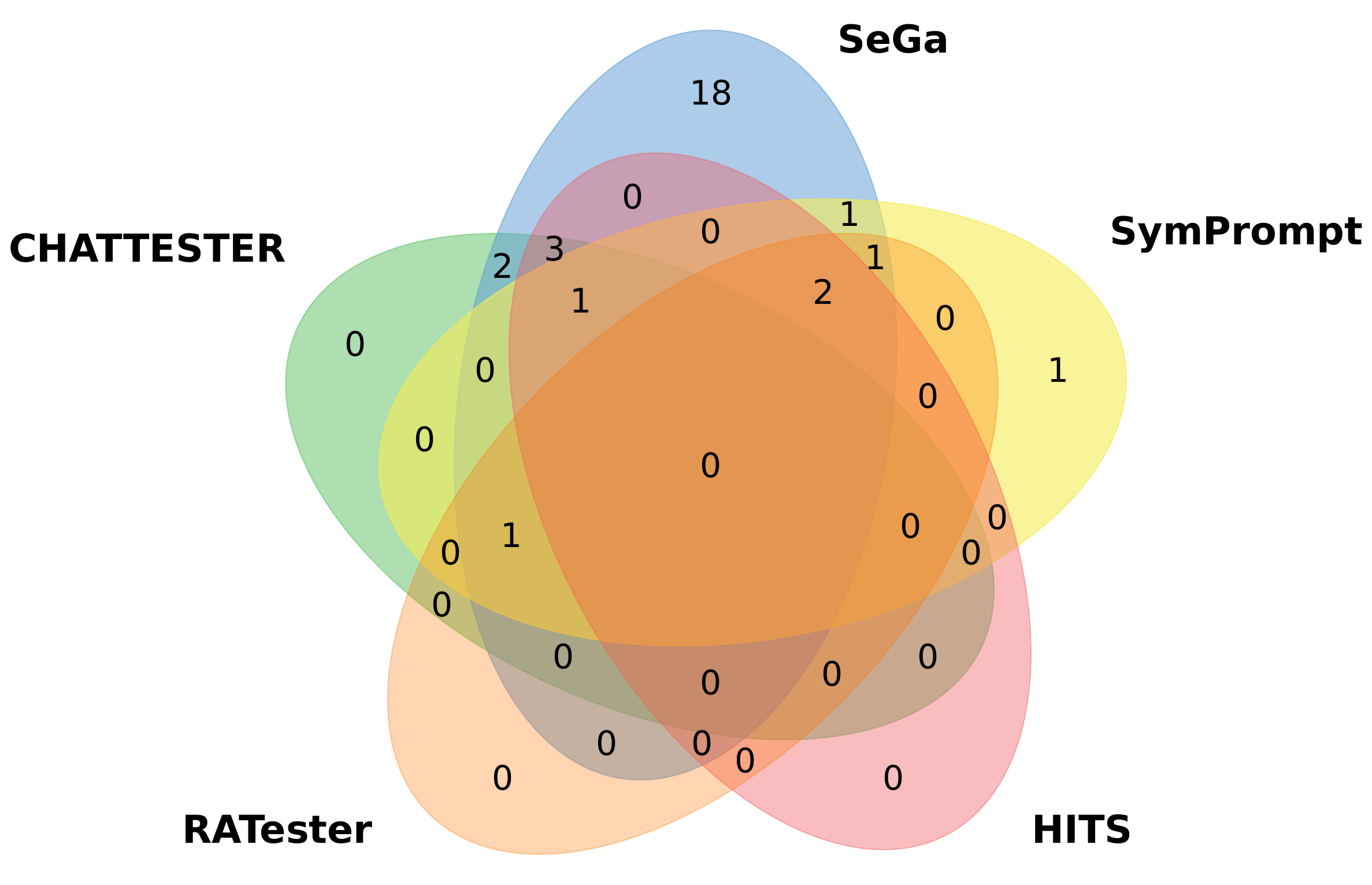}
  \caption{Overlap of bug detection}
  \label{fig:rq1_venn}
  \vspace{-2em}
\end{wrapfigure}

Figure~\ref{fig:rq1_venn} illustrates the overlap of bugs detected by different techniques. From the figure, \tool{} identifies 29 bugs in total, while \tester{}, \sym{}, \hits{}, and \rat{} detect 7, 7, 6, and 4 bugs, respectively. Notably, \tool{} covers nearly all bugs found by the other techniques (missing only one uniquely detected by \sym{}) and also discovers the largest number of bugs that no baseline can find. This highlights its strong practical advantage in real-world bug discovery.
The weak performance of the baselines stems from their code-centric nature. 
They infer intent from the existing implementation via intent summarization (\tester{}), path/fragment decomposition (\sym{}, \hits{}), or local context retrieval (\rat{}), which biases generated tests to conform to current code and rarely exposes \bugtype{}s. 
In contrast, \tech{} incorporates PRD-derived semantics to construct business scenarios that can violate the implementation, enabling it to uncover high-level \bugtype{}s that code-centric methods often miss.

We further analyzed the false positives and identified several recurring causes.
First, some false positives arise when the model hypothesizes implausible failure scenarios that are unlikely in deployed production systems.
For example, \tech{} may flag a method as buggy by assuming downstream services return malformed responses or that a database intermittently violates its consistency guarantees, even though such failures are prevented by well-tested infrastructure and operational safeguards.
Second, false positives can be triggered by ambiguities, inconsistencies, or missing details in PRDs.
For instance, when a PRD does not specify whether a boundary case (e.g., empty input, legacy state, or default configuration) should be rejected or tolerated, \tech{} may infer a stricter behavior than intended and report a violation.
Finally, most false positives stem from context-dependent design choices: code patterns that appear risky in isolation may be intentionally adopted under team-specific engineering conventions.
For example, we observed cases where \tech{} flagged transient plaintext password handling as a security bug.
In a database-backed authentication module, persisting plaintext passwords violates the module’s responsibility to enforce credential protection and constitutes a genuine bug.
In contrast, in a user interaction layer, plaintext passwords are handled only transiently for validation and request forwarding, with hashing and secure storage intentionally delegated to downstream services, therefore causing a false positive.
This illustrates that accurate \bugtype{} detection depends not only on explicit requirements, but also on module responsibilities and project-specific practices that may not be fully captured in PRDs.

\begin{tcolorbox}[colback=gray!5]
\textbf{RQ1 Summary}: \tool{} outperforms all baselines in \bugtype{} detection, achieving the highest precision, recall, and F1-score. 
That is, it not only detects the most \bugtype{}s, including many unique ones, but also maintains low false positive rates.
\end{tcolorbox}

\subsection{RQ2: Ablation Study}
\subsubsection{Process}
In this RQ, we study how \tool{}’s key components contribute to effectiveness: \yc{semantic knowledge construction, semantic context retrieval, and compilation repair}. 
Since semantic context retrieval is central to \tool{}, we further ablate its two sub-steps: relevant functionality retrieval and scenario derivation.
In total, we construct \yc{four} ablated variants of \tool{}.
\begin{itemize}
\item \textbf{\nobase{}} removes semantic knowledge construction. The pipeline uses raw requirement text instead of a structured knowledge base.
\item \textbf{\nofunc{}} removes relevant functionality retrieval. All functionality entries in the knowledge base are passed to the next stage.
\item \textbf{\noscenario{}} removes scenario derivation. Retrieved functionalities are used as-is without being refined into scenarios with explicit preconditions, actions, and expected outcomes.
\item \yc{\textbf{\norepair{}} removes the standalone compilation repair. The agent iteratively fix compilation errors itself.}
\end{itemize}
\noindent
We apply all variants to all focal methods using the same setup as RQ1.

\subsubsection{Results}

Table~\ref{tab:rq2} shows that removing any component of \tech{} degrades performance, indicating that each component contributes to \bugtype{} detection.
\nobase{} performs worst overall. 
Without semantic knowledge construction, it feeds raw requirement documents into downstream stages. 
This distracts the agent and weakens the useful signals for testing. 
As shown in Table~\ref{tab:rq2}, \nobase{} produces consistently low F1-scores on Subject-1 (0.24), Subject-2 (0.32), and Subject-3 (0.27), with particularly low precision (0.23 on Subject-1 and 0.24 on Subject-3). 
This confirms that simply providing full PRDs is insufficient; the knowledge base is necessary to distill and organize requirement information into usable functionality entries.
\nofunc{} shows relatively higher recall but low precision. Although it does not retrieve relevant functionality entries, the full functionality set still contains rich information, so the model can sometimes pick out relevant fragments and detect bugs. However, without relevance filtering, it is easily misled by unrelated functionalities, leading to many false positives. 
This indicates that functionality retrieval is essential to focus the agent on the subset of functionalities that matter for the focal method.
\noscenario{} achieves relatively higher precision but much lower recall. 
It retrieves relevant functionality entries but does not derive scenario-level guidance, leaving the generation agent with coarse descriptions only. 
This tends to reduce false positives because, without scenario-level guidance, the generator usually produces generic tests with weaker or incomplete assertions.
However, without explicit preconditions, actions, and outcomes, the agent struggles to construct inputs that trigger subtle, context-dependent bugs.
For example, on Subject-3, \noscenario{} reaches 0.59 precision but only 0.35 recall, showing that many true \bugtype{}s are missed without scenario derivation.
\yc{
\norepair{} performs better than \nobase{}, but still falls behind other variants and the full \tech{}. A major reason is that the agent spends many iterations handling compilation errors by searching for missing context, rewriting tests, and repeatedly re-triggering new errors, until it reaches the iteration budget. 
This result confirms the benefit of standalone compilation repair.
}

Overall, the full \tool{} performs the best in terms of all metrics across all subjects. The results suggest that effective \bugtype{} detection requires all components to work together: knowledge construction to reduce PRD noise, functionality retrieval to ensure method relevance, scenario derivation to concretize requirements into test-ready guidance, and compilation repair to make generated tests executable.
\begin{tcolorbox}[colback=gray!5]
\textbf{RQ2 Summary}: All components of \tech{} contributes to the overall effectiveness. 
Together, they enable reliable and effective \bugtype{} detection.
\end{tcolorbox}
\begin{table}[t]
\renewcommand{\arraystretch}{1}
\caption{Comparison between \tool{} and its variants}
\centering
\label{tab:rq2}
\resizebox{0.99\linewidth}{!}
{
\begin{threeparttable}
\begin{tabular}{l|ccc|ccc|ccc|ccc}
\toprule
\multicolumn{1}{c|}{\multirow{2}{*}{\textbf{method}}} & \multicolumn{3}{c|}{\textbf{Subject-1}} & \multicolumn{3}{c|}{\textbf{Subject-2}} & \multicolumn{3}{c|}{\textbf{Subject-3}} & \multicolumn{3}{c}{\textbf{Subject-4}} \\
\multicolumn{1}{c|}{} & \multicolumn{1}{c}{\textbf{Precision}} & \multicolumn{1}{c}{\textbf{Recall}} & \multicolumn{1}{c|}{\textbf{F1}} & \multicolumn{1}{c}{\textbf{Precision}} & \multicolumn{1}{c}{\textbf{Recall}} & \multicolumn{1}{c|}{\textbf{F1}} & \multicolumn{1}{c}{\textbf{Precision}} & \multicolumn{1}{c}{\textbf{Recall}} & \multicolumn{1}{c|}{\textbf{F1}} & \multicolumn{1}{c}{\textbf{Precision}} & \multicolumn{1}{c}{\textbf{Recall}} & \multicolumn{1}{c}{\textbf{F1}} \\
\midrule
\nobase{}   & 0.23                                & 0.25                                & 0.24                                & 0.30                        & 0.33                       & 0.32                       & 0.24                       & 0.30                        & 0.27                       & 1.00                              & 1.00                              & 1.00                              \\
\nofunc{}    & 0.31                          & 0.33                          & 0.32   & 0.38                                   & 0.56                                   & 0.45                                   & 0.35                          & 0.35                          & 0.35                                                & 0.50                              & 0.50                              & 0.50                              \\
\noscenario{} & 0.40                                 & 0.17                                & 0.24                                & 0.57                       & 0.44                       & 0.50                        & 0.59                       & 0.35                       & 0.44                       & 1.00                              & 1.00                              & 1.00                             \\
\norepair{}     & 0.50                                   & 0.17                                   & 0.25                                   & 0.38                          & 0.33                          & 0.35                          & 0.35                          & 0.32                          & 0.34                          & 1.00                              & 1.00                              & 1.00                              \\
\midrule
\tech{} & \textbf{0.71} & \textbf{0.42} & \textbf{0.53} & \textbf{0.86} & \textbf{0.67} & \textbf{0.75} & \textbf{0.67} & \textbf{0.43} & \textbf{0.52} & \textbf{1.00} & \textbf{1.00} & \textbf{1.00} \\
\bottomrule
\end{tabular}
\end{threeparttable}
 }
\end{table}

\subsection{RQ3: Detecting New Business Bugs}
\subsubsection{Process}
In this RQ, we evaluate \tool{}’s ability to discover \bugtype{}s under real development workflows.
As introduced in Section~\ref{sec:subject}, we deployed \tool{} on \deploynum{} internal production repositories from multiple business lines, covering diverse application domains.
Specifically, \tool{} was integrated into the CI/CD pipeline and automatically triggered when a merge request (MR) was opened. 
For each MR, we identified all methods modified or newly introduced and treated them as focal methods. 
The semantic knowledge base is constructed in advance from the associated requirement documents. 
For each focal method, \tool{} retrieves relevant functionality entries from the knowledge base, derives business scenarios by jointly reasoning over the retrieved functionalities and the focal method’s implementation, and then generates unit tests guided by these scenarios.

Generated tests were executed against the MR version of the code. 
Test failures were reported back to the developer who opened the MR.
Developers then inspected the reported issue. 
A reported issue was counted as a true \bugtype{} only if the developer confirmed that it reflected an unintended deviation from the requirements. 
Importantly, these MRs had already passed the projects’ existing test suites and developers’ local validation before submission, reflecting standard engineering practice.
This process ensures that the detected bugs are both previously unknown and practically relevant to ongoing development.
We ran the deployment continuously for one week to evaluate \tool{} in isolation,
and we report only the results from this period to avoid confounding effects from the company’s existing internal testing tools, into which \tool{} was later integrated.

\subsubsection{Results}

\begin{figure*}[t]
  \centering
\includegraphics[width=\linewidth]{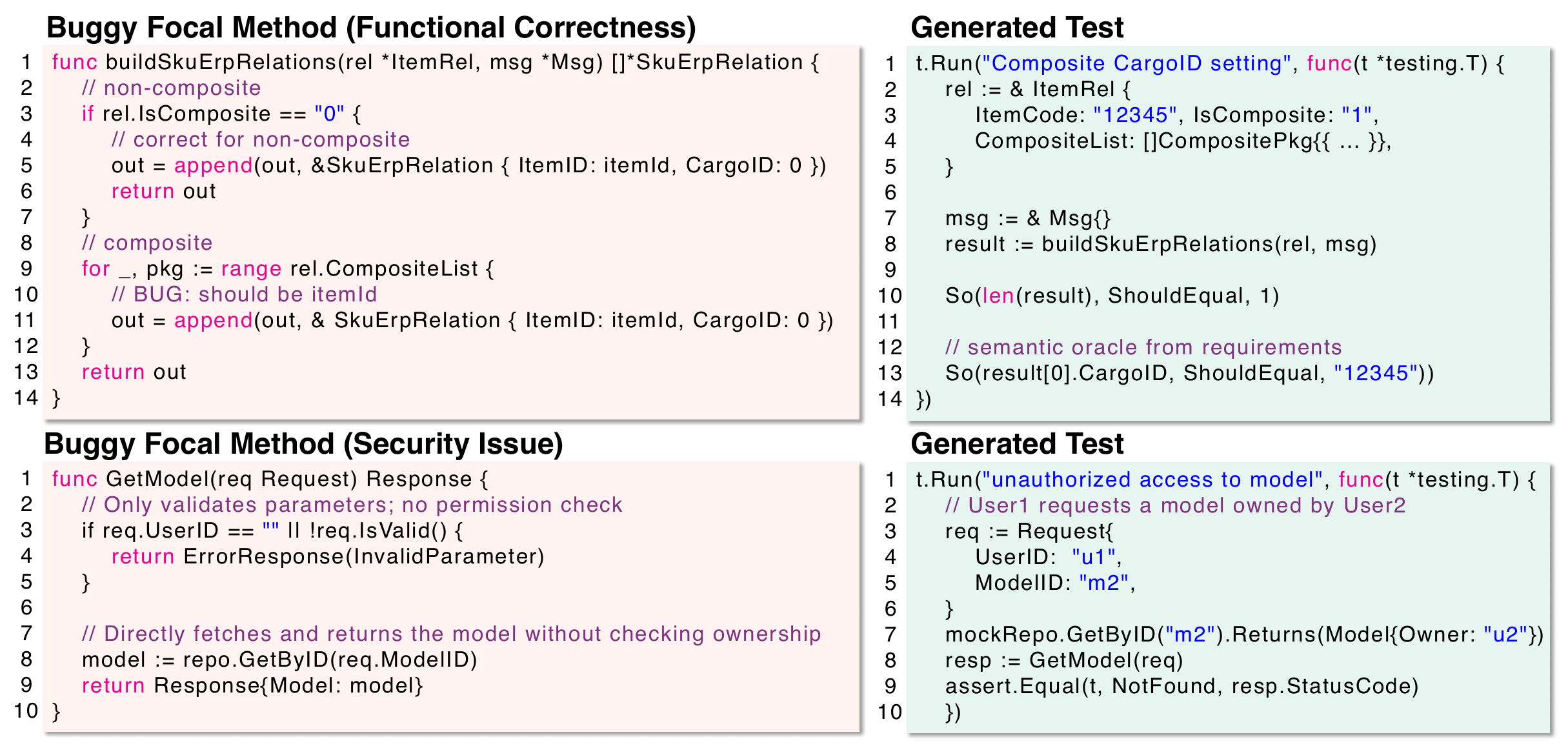}
  \caption{Simplified and desensitized versions of two previously unknown \bugtype{}s}
  \label{fig:bug_example}
 \vspace{-2em}
\end{figure*}

Across the production repositories, \tool{} successfully detected \fixed{} previously unknown \bugtype{}s, all of which have been fixed by developers, while none of the baseline techniques reported any of these issues. 
\yc{
We further analyzed the detected bugs and found that they span the major bug categories introduced earlier.
The corresponding buggy code changes involve \textit{91.35} lines on average and often implement the correct control flow yet violate documented requirements due to mismatched conditions, states, or cross-component assumptions.
This result demonstrates that semantics-guided test generation can expose \bugtype{}s that remain largely invisible to conventional code-centric approaches in real CI/CD settings.
}
\yc{
Regarding false positives, \tool{} occasionally reports behaviors that appear suspicious when evaluated solely against requirement text but are intentional under project-specific conventions (\textit{~30}\% of the total reports).
As discussed earlier, such cases typically stem from incomplete or ambiguous requirement documentation, implausible failure assumptions about downstream components, or context-dependent design decisions that are not visible to the tool.
Importantly, developers were able to quickly dismiss these reports, indicating that while false positives exist, they are generally explainable and do not overwhelm the practical usefulness of the detected bugs.
}

Figure~\ref{fig:bug_example} presents two representative examples from different bug categories: a functional correctness bug and a security bug. 
In both cases, the documented requirements specify expected software behavior that is not explicit in the code, making the bugs difficult for code-centric unit test generation techniques to expose.
Regarding the functional correctness bug, the requirement document states that, for composite items, the parent item identifier must be stored in the \texttt{CargoID} field so downstream systems can trace bundle membership, while for non-composite items this field should remain zero. 
However, the focal method
assigns \texttt{CargoID} to zero in both branches (Line~5 and Line~11). 
\tool{} retrieves the relevant functionality, and derives a business scenario for processing a composite item.
The generated test asserts that \texttt{CargoID} must equal the parent identifier under composite conditions and fails at runtime, revealing the bug.
Regarding the security bug, the requirement document specifies that a user may only retrieve models they own, requests for models owned by other users should be denied. 
In \texttt{GetModel}, the implementation validates parameters (Line~3) but performs no ownership check before querying the repository (Line~8), and thus returns another user’s model. Similarly, \tool{} derives a business scenario in which a user requests a model they do not own; the generated test asserts rejection and fails, exposing the bug.

\begin{tcolorbox}[colback=gray!5]
\textbf{RQ3 Summary}: \tool{} demonstrates strong practical effectiveness in detecting previously unknown \bugtype{}s in actively maintained enterprise projects.
\end{tcolorbox}

\section{Lessons Learned}
\label{sec:discussion}

\textbf{Business logic bug detection requires semantics, but relevance matters more than volume.}
Our study shows that many high-impact business logic bugs cannot be detected from source code alone, because they arise from violations of intended business behavior captured outside the codebase, typically in PRDs.
However, increasing semantic context by indiscriminately feeding more PRD content does not monotonically improve effectiveness.
In \tech{}, even when functionality entries are semantically correct, incorporating those that are weakly related to the focal method often degrades performance, because irrelevant functionalities distort scenario derivation. The derived business scenarios may over-approximate expected behavior, producing tests that are misaligned with the intended scope of the focal method.
Therefore, future semantics-driven testing approaches should prioritize relevance-aware semantic selection over maximizing semantic coverage. 
Improving the ability to identify and retain only that semantic information that are tightly coupled to the focal method is likely more impactful than enriching the overall amount.

\smallskip
\noindent
\textbf{False positives often capture a gap between best practice and enterprise practice.}
During industrial deployment, we observed that a substantial portion of false positives rejected by developers as non-bugs arose from implicit organizational decisions, such as responsibility boundaries between modules, legacy compatibility constraints, or intentionally deferred validation logic, which were not encoded in the PRDs.
As a result, test failures often conflicted with how the software was expected to behave within its organizational and architectural context.
That is, reducing false positives in practice requires more than improving semantic extraction from PRDs. Semantics-driven testing tools must account for project-specific conventions and responsibility boundaries that shape how requirements are interpreted. Supporting mechanisms for incorporating such knowledge is essential for improving developer acceptance.

\smallskip
\noindent
\textbf{Richer semantics and feedback loops are needed to move beyond PRD-to-scenario.} 
\tool{} shows that turning PRDs into business scenarios can guide unit testing for \bugtype{} detection, but it is only one point in a large design space. 
Many choices about how to mine and use business semantics remain open. 
One direction is to explore richer representations of business semantics beyond our current DSL, such as explicitly modeling state transitions, data lifecycles, or capturing trace links among requirements and code elements to make reasoning easier. 
Moreover, future techniques would benefit from tighter feedback loops. Instead of treating requirements as static input, they could use test execution results, developer reviews, and requirement updates to continuously refine retrieved semantics and to support long-lived test suites that evolve with changing business semantics and reducing false positives over time.

\smallskip
\noindent
\textbf{Code-change-aware deployment is essential for sustainable semantics-driven testing.}
Our deployment experience suggests that, while semantics-driven testing can be applied at the repository level, its practical value in CI/CD pipelines depends on being deployed in a change-aware manner.
Business semantics are global and stable, but the review cost is incurred per code change.
Prioritizing focal methods affected by a change and generating a limited number of tests by targeting only relevant business scenarios per method allows semantics-driven testing to scale without overwhelming developers.
This strategy preserves the benefits of semantic knowledge while ensuring that reported issues remain actionable and aligned with developers’ responsibilities.

\smallskip
\noindent
\textbf{Co-evolution of requirements and code.}
In long-lived projects, requirement documents (e.g., PRDs) are often created at a specific time and evolve much more slowly than the codebase. As features are incrementally modified or extended, the code may diverge from the original specification, leading to inconsistencies that do not necessarily indicate bugs but rather outdated or incomplete documentation. This suggests the need to treat requirements and code as co-evolving artifacts. Instead of assuming requirements as a fixed oracle, future systems should explicitly model and support their synchronization, e.g., by detecting documentation drift, triggering requirement updates, or jointly evolving specifications alongside code changes.

\smallskip
\noindent
\textbf{Oracle hallucination.} When some scenarios are underspecified (e.g., constraints on edge cases, boundary conditions, or exceptional behaviors), the system may over-interpret the intended logic and generate unrealistic or overly strict test scenarios. We observed that, in such cases, the generated oracles sometimes reflect plausible but unintended behaviors, leading to false positives. This issue is particularly pronounced when natural language descriptions leave room for multiple valid interpretations. That is, there exists a safety-utility trade-off in requirement-driven test generation: insufficient constraints can lead to “over-testing” beyond the intended business logic. Effective systems should bound their reasoning scope or incorporate uncertainty-awareness to avoid over-constraining expected behavior.

\smallskip
\noindent
\textbf{Cross-repository requirements.} In modern software systems, especially those based on microservices or distributed architectures, requirements may span multiple repositories or services. We observed cases where correct behavior depends on interactions across services, and partial visibility can lead to incomplete or misleading interpretations of requirements. This suggests that supporting cross-repository requirement reasoning is critical. This remains an important direction for future work, potentially requiring integration across multiple codebases.

\section{Threats to Validity}
The \textbf{external threat} primarily stems from the generalizability.
Our study uses repositories from a single company and focuses on one programming language (Go). 
We evaluate industrial repositories because \tool{} relies on system requirements (e.g., PRDs) to obtain business semantics, which are rarely available or sufficiently detailed in open-source projects.
Moreover, common open-source unit-test benchmarks (e.g., Defects4J~\cite{defects4j}, BugsInPy~\cite{bugsinpy}) largely consist of libraries with limited business logic and thus are a poor proxy for \tool{}'s target setting. 
To mitigate this threat, we evaluate \tool{} on four production Go repositories from distinct business lines and production types and additionally deploy it on \deploynum{} independent large-scale repositories maintained by different teams, where it uncovered \fixed{} previously unknown \bugtype{}s. 
These results provide evidence that \tool{} generalizes across projects, domains, and development contexts, partially mitigating this threat.
Regarding agile processes where there are only user requirements and no or few system requirements, user requirements can be lightly normalized and grounded with code context to reduce abstraction. 
In scenarios where explicit requirements are unavailable, \tool{} could be further extended, as its core idea is to progressively decompose natural language specifications, establish their correspondence with code, and guide test generation. Alternative artifacts, such as issue descriptions or developer comments, may serve as partial substitutes. 
In the future, we plan to validate \tool{} on additional projects, programming languages, and scenarios.

The \textbf{internal threats} primarily stem from potential implementation errors in \tool{} or in the compared techniques.
To reduce this risk, we directly use the official implementation of \rat{}~\cite{rat}. For \tester{}, \sym{}, and \hits{}, we adapt their publicly available implementations to Go by strictly following the descriptions in their respective papers and validate their behavior on representative examples. 
The implementation of \tool{} itself was carefully reviewed and tested.

The \textbf{construct threats} include LLM randomness, potential data leakage, and the suitability of evaluation metrics.
To control randomness, we set the LLM temperature to zero for all techniques, following established practice in prior work~\cite{ase_empirical,fuzz4all,dan}.
To address data leakage concerns, all evaluated repositories are proprietary industrial codebases that are not publicly accessible and were confirmed by our partner to be excluded from any LLM training data. 
Regarding metrics, we used widely-used metrics, Precision, Recall, F1-score, to measure effectiveness. 
\yc{
We also examined the overhead of \tool{}. On average, \tech{} takes 236.85 seconds per focal method, with \tester{} requiring 139.91 seconds, \sym{} 187.11 seconds, and \hits{} 134.74 seconds per method. Although these compared techniques incur realatively fewer computational costs, they detect substantially fewer \bugtype{}s than \tool{}. Given the high real-world cost of \bugtype{}s, the overhead of \tool{} is justified by its improved effectiveness.
}

\section{Related Work}
\smallskip
\noindent
\textbf{LLM-based Unit Test Generation.}
A number of techniques explore LLM-based unit test generation by prompting models with source code plus additional context, often aided by lightweight program analysis, for both popular languages (e.g., Java and Python)~\cite{chattester, symprompt, wang2024hits, JUnitGenie, agonetest, wiseut, clast, ase_empirical, context, dl_study} and less-studied languages (e.g., C, Rust, and Go)~\cite{strut, rug, palm, rat}. 
Beyond direct prompting, other work integrates LLMs with traditional testing techniques~\cite{codamosa, telpa}, or improves test generation by training models~\cite{zhang2025less, shin2023domain}.
Recently, IntentionTest~\cite{intentiontest} explores intention-driven test generation by providing LLMs with explicit validation intentions.
It adopts a retrieval-and-edit strategy that reuses existing project tests and incorporates project-specific code context to generate tests aligned with the given validation intention.

A few recent studies move toward bug-oriented test generation.
Xin et al.~\cite{auger} introduce an attention-based mechanism to identify defective Java methods and guide LLMs to generate bug-revealing tests.
UnitCon~\cite{unitcon} leverages static analysis to estimate test semantics and prioritize candidates likely to trigger Java runtime exceptions.
RTED~\cite{rted} combines step-wise type constraint analysis with reflective validation to detect Python type errors.

However, these approaches either target specific low-level bug classes (e.g., runtime exceptions or type errors) or depend on supervised learning with large amounts of fine-grained bug annotations (e.g., faulty-line labels), which is costly and hard to scale. Such annotations are particularly difficult to obtain for \bugtype{}s, where failures stem from requirements that are not explicitly reflected in code structure. \textit{In contrast, \tech{} incorporates business semantics extracted from PRDs to guide LLM-based unit test generation. By complementing code context with semantic context, it enables systematic discovery of \bugtype{}s that existing techniques often miss.}

\smallskip
\noindent
\textbf{Requirement-Based Testing.}
A substantial body of research has explored generating test cases from requirement specifications, commonly referred to as requirements-based test generation (RBTG).
A recent survey by Yang et al.~\cite{yang2025requirements} categorizes RBTG techniques by requirement types, modeling formalisms, and test generation strategies, and identifies natural-language requirements as a persistent challenge.
Early RBTG approaches primarily transformed formal requirement models, such as state machines or logical specifications, into test sequences~\cite{rayadurgam2001test,sharma2014automated}. 
Subsequent work leveraged use cases or structured SRS documents to derive acceptance- or system-level tests~\cite{wang2020automatic,raamesh2010reliable}. 
While effective within their scope, these techniques depend on formalized or semi-structured artifacts and scale poorly to the raw, free-form PRDs that dominate industrial practice.
More recent efforts apply NLP pipelines and LLM-assisted techniques to generate tests directly from textual requirements~\cite{nlp_pipeline,gpt4test2024}. 
Despite their promise, such approaches typically produce natural-language test cases, offering limited alignment with specific code units and insufficient precision to expose subtle business logic bugs.
Complementary work on model-based testing frameworks (e.g., UML-derived models) and formal requirement elicitation tools (e.g., FRET) enable more precise and traceable test generation~\cite{stocks1993test,frettool}, but typically rely on manually constructed models.
\textit{In contrast, \tech{} targets unit test generation from raw PRDs. By constructing a semantic knowledge base of functionalities, retrieving those relevant to a focal method, and deriving business scenarios with semantic constraints to guide test generation, \tool{} generates unit tests aligned with requirements and enables automated detection of \bugtype{}s at the unit level.}

\section{Conclusion}
In this paper, we present \tech{}, a semantics-driven unit test generation technique for effectively detecting business logic bugs. 
\tech{} builds a structured semantic knowledge base from product requirement documents, consisting of functionality entries that group related requirements under a common business intent.
Given a focal method, \tech{} retrieves the relevant functionality entry and derives fine-grained business scenarios (with preconditions, triggering actions, expected outcomes, and semantic constraints) to guide LLM-based test generation.
Our evaluation on four industrial Go projects shows that \tech{} can detect 22$\sim$25 more real world business logic bugs than state of the art LLM based techniques, including \tester{}, \sym{}, \hits{}, and \rat{}, and improves precision by 26.9\%$\sim$34.6\%.
In addition, deployment across \deploynum{} production repositories uncovers \fixed{} previously unknown business logic bugs confirmed and fixed by developers.

\section*{Data Availability} 
Due to confidentiality agreements with our industrial partner, we cannot release the evaluated projects and the full source code of \tech{}. However, to support transparency and reproducibility, we provide the core component of \tool{} (i.e., the semantic reasoning agent).
We also provide representative, desensitized examples. Please find them on our project homepage~\cite{homepage}.

\section*{Acknowledgment}
We thank all the ISSTA anonymous reviewers for their valuable comments.
This work was supported by the National Natural Science Foundation of China (Grant Nos. 62322208, 62232001).

\normalem
\bibliographystyle{ACM-Reference-Format}
\bibliography{sections/10_ref}

\end{document}